\documentclass[aps,pra,showpacs,twocolumn,amsmath,amssymb,superscriptaddress,footinbib]{revtex4}

\usepackage{graphicx}
\usepackage{dcolumn}
\usepackage{bm}
\usepackage{subfig}

\usepackage[english]{babel}
\usepackage{latexsym}
\usepackage{graphicx}
\usepackage{epsfig}
\usepackage{amsfonts}
\usepackage{amssymb}
\usepackage{amsmath}

\begin{document}

\preprint{APS/123-QED}

\title{A diabatic representation of the two lowest electronic states of Li$_3$}

\author{Elham Nour Ghassemi}
\affiliation{Department of Physics,
Stockholm University, AlbaNova University Center, SE-106 91 Stockholm,
Sweden}
\author{Jonas Larson}
\affiliation{Department of Physics,
Stockholm University, AlbaNova University Center, SE-106 91 Stockholm,
Sweden}
\affiliation{Institut f\"{u}r Theoretische Physik, Universit\"{a}t zu
  K\"{o}ln, K\"{o}ln, De-50937, Germany}
\author{{\AA}sa Larson}
 \email{aasal@fysik.su.se}
\affiliation{Department of Physics,
Stockholm University, AlbaNova University Center, SE-106 91 Stockholm,
Sweden}

\date{\today}

\begin{abstract}
Using the Multi-Reference Configuration Interaction method, the adiabatic potential energy surfaces of Li$_3$ are computed. The two lowest electronic states are bound and exhibit a conical intersection. By fitting the calculated potential energy surfaces to the cubic $E\otimes\varepsilon$ Jahn-Teller model we extract the effective Jahn-Teller parameters corresponding to Li$_3$. These are used to set up the transformation matrix which transforms from the adiabatic to a diabatic representation. This diabatization method gives a Hamiltonian for Li$_3$ which is free from singular adiabatic couplings and should be accurate for large internuclear distances, and it thereby allows for bound dynamics in the vicinity of the conical intersection to be explored.

\end{abstract}

\maketitle

\section{Introduction}
In molecular systems with three or more atoms, degeneracy occurs in highly symmetrical nuclear configurations. These degeneracies typically lead to a breakdown of the Born-Oppenheimer description and hence greatly influence the molecular dynamics. The simplest example of such a situation is given by the $E\otimes\varepsilon$ Jahn-Teller (JT) effect \cite{Jahn-Teller-37} with a conical intersection involving the two-fold degenerate electronic states {\it E} with the doubly degenerate nuclear vibration modes $\varepsilon$. Alkali metal trimers X$_3$ with a conical intersections at the geometric $D_{3h}$ symmetries are special cases of the $E\otimes\varepsilon$ JT systems~\cite{Mead-85-1}. As a result, in recent years small metal clusters have received great attention as the subject of many experimental  \cite{hermann78,thompson82,woste86,broyer86,broyer87,broyer88,broyer89,broyer89-2,demtroder88,broyer90} and theoretical \cite{martin78,schumacher78,schumacher82,martin83,Mead-85-1, Mead-85-2,andreoni88,martin89, meyer99} publications. These systems are especially convenient for experimental studies due to their possibility of vibrational excitation in the visible or near infrared regions \cite{Varandas-98}. The two lowest potential energy surfaces for Li$_3$ host a conical intersection at $D_{3h}$ symmetry \cite{Mead-85-1,Mead-85-2}.  Moreover, the lower states are bound which makes it possible to follow the dynamics of the system in the vicinity of the conical intersection. This is an interesting regime for many reasons, for example because the JT theorem tells us that the electronic groundstate ($1^2E$) at the highly symmetric and degenerate nuclear configuration is dynamically unstable, and the system will therefore distort into a state with lower symmetry ($1^2B_2$) at a non-symmetric nuclear configuration. 

All these compelling properties of Li$_3$ together with its simplicity due to the few number of electrons has made it a well studied system in terms of the dynamical JT effect. The potential energy surfaces of the lower electronic states of Li$_3$ have been studied before using quantum chemistry. Already in 1978, Gerber \textit{et al.} \cite{schumacher78} computed the lowest adiabatic potential energy surface as a function of the normal mode coordinates using the coupled electron pair approximation (CEPA). The lowest adiabatic potential energy surface has also been calculated by applying density functional theory with a pseudopotential approximation~\cite{martin83}. Using the Multi-Reference Configuration Interaction (MRCI) method, Ehara \textit{et al.} \cite{ehara99} calculated the full three-dimensional potential energy surfaces of the first and second adiabatic states. They continued by performing nuclear wave packet propagations on the lowest adiabatic (i.e. uncoupled) potential energy surface for modeling of the stimulated emission pumping spectrum. Finally, Thompson \textit{et al.} \cite{Mead-85-2} constructed an analytical representation of the two lowest \textit{ab initio} potential energy surfaces as a function of the internuclear coordinates. 

For Li$_3$, the adiabatic potential energy surface of the ground electronic state exhibits three global minima separated by low potential barriers. The shallowness of the minima corresponding to the isosceles configurations makes the potential energy surface to be relatively flat. Tunneling from one minima to another minima causes pseudorotation of the trimer. In order to correctly explain and simulate effects like this, the full Hamiltonian is needed including couplings between the adiabatic or diabatic states. Instead of describing the molecular system in its adiabatic representation with non-adiabatic coupling elements that become singular at the conical intersection, a diabatic representation is here considered. Normally, in the case of a JT effect, the Hamiltonian is often approximated with a diabatic JT Hamiltonian. This is based on a Taylor expansion of the energy surfaces and couplings around the conical intersection. 
An alternative approach is not to use the JT Hamiltonian directly, but to exploit the fitted JT parameters to construct the unitary transformation matrix that transforms the states from the adiabatic to the diabatic representations. Then the {\it ab initio} adiabatic potential energy surfaces are transformed to {\it quasidiabatic} potential energy surfaces and couplings. The advantage of this approach is that a reliable description of the system is obtained even far from the conical intersection, where the JT model is no longer valid. This diabatization scheme that we use was originally suggested by Thiel and K\"{o}ppel~\cite{koppel99}. 
   
In the present work, the three-dimensional adiabatic potential energy surfaces of the two lowest electronic states of Li$_3$ are computed using the MRCI method. The quality of these \textit{ab initio} calculations is investigated by first computing the potential energy curves of the Li$_2$ diatomic molecule. The adiabatic potential energy surfaces are then fitted to the eigenvalues of the potential part of the JT Hamiltonian including terms up to third order. The parameters are compared with data obtained in former studies. With the method of \cite{koppel99}, the adiabatic potentials are transformed to a diabatic representation. This diabatic potential energy matrix describes the system not only close to the conical intersection, but also further away from it. The diabatic potentials of Li$_3$ computed here can be employed to study the influence of the conical intersection upon either stationary states or molecular dynamics. 

The outline of the paper is the following. In the proceeding section we give a brief discussion about adiabatic vs. diabatic representations and the normal mode coordinates of the Li$_3$ trimer are presented. This section also presents some general theory of the $E\otimes\varepsilon$ JT model, for example giving the Hamiltonian up to cubic order and its corresponding adiabatic-to-diabatic transformation matrix. In section~\ref{sec3} we present the results from the quantum chemistry calculations of the adiabatic potential energy surfaces. In the following section~\ref{sec:diab} the diabatization method is explained and the obtained JT parameters are given together with an analysis of the diabatic Hamiltonian. Finally we end with concluding remarks in section~\ref{sec5}.

\section{Theory}\label{sec2}
For later purposes, we here discuss the adiabatic and diabatic representations as well as the normal mode coordinates of Li$_3$. We then continue by introducing the JT Hamiltonian and its adiabatic-to-diabatic transformation matrix. With the general theory presented in this section, the method of diabatization (which is a main objective of the present work) is rather straightforward. Throughout the paper we will use atomic units.  

\subsection{Adiabatic and diabatic representations}
In general, the non-relativistic molecular Hamiltonian is given by the sum of the nuclear kinetic energy operator $T$ and the electronic Hamiltonian $H_{el}$, i.e. $H=T+H_{el}$. The adiabatic electronic states are defined as the states that diagonalize the electronic Hamiltonian $H_{el}$ at fixed nuclear coordinates. Using these states as a basis for expansion of the total molecular wave function leads to a coupled nuclear Schr\"{o}dinger equation, where the motions on the different adiabatic potential energy surfaces are coupled by non-adiabatic coupling terms originating from off-diagonal matrix elements of the nuclear kinetic energy operator. These terms blow up when the energy gap between the adiabatic potential energy surfaces becomes small. Such large matrix elements naturally cause problems when running any numerical study on the system. Note that the adiabatic electronic states are only defined up to an overall phase factor that could depend on the internuclear coordinates. This phase ambiguity is a manifestation of the underlying gauge structure hidden in Born-Oppenheimer dynamics~\cite{baer06}. We will meet this gauge freedom once more when we discuss the adiabatic-to-diabatic transformation matrix. 

At a conical intersection, the gap between two adiabatic potential energy surfaces closes and the two states become degenerate. The non-adiabatic couplings may then diverge. This is the essence of the $E \otimes \varepsilon$ JT effect. As already announced, in situations where non-adiabatic corrections play a major role it can be advantageous to change basis to a diabatic representation via an unitary adiabatic-to-diabatic transformation. A strict diabatization requires that all non-adiabatic coupling elements completely vanish. Since the diabatic states are not eigenstates to the electronic Hamiltonian, in the diabatic representation the potential matrix will no longer be diagonal. The adiabatic and diabatic potential energy matrices are related by
\begin{equation}
\label{eq:a2d}
{\bf U}^a={\bf S}^{\dagger}{\bf U}^d{\bf S},
\end{equation}
where ${\bf S}$ is the adiabatic-to-diabatic transformation matrix. It is clear that in order to perform a strict diabatization one would need the knowledge of the non-adiabatic couplings among the adiabatic states~\cite{maed82,baer06}. To circumvent such difficulties, a common approach is to use {\it quasidiabatic} states in which the non-adiabatic couplings do not completely disappear. This is the case for the JT Hamiltonian, where the leading terms of the diabatic potential matrix are included. 

\subsection{Normal mode coordinates}
\begin{figure}[h]
	\centering{
	\includegraphics[width=8cm]{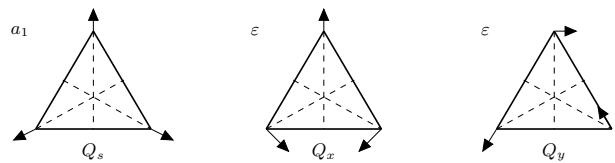}}
	\caption{The three normal modes of the  Li$_3$ molecule; stretching $Q_s$, bending $Q_x$, and asymmetric stretching $Q_y$. }	
	\label{fig:imd3h}
\end{figure}

The three normal mode coordinates of Li$_3$ are associated with the symmetric stretching, bending and asymmetric stretching motions. They are denoted by $Q_s$, $Q_x$ and $Q_y$, and are illustrated in Fig.~\ref{fig:imd3h}. 

In $D_{3h}$ symmetry, the $Q_s$ motion belongs to the $a_1$ irreducible representation, while $Q_x$ and $Q_y$ belong to the degenerate $\varepsilon$ mode. If the Cartesian coordinates of the atoms in the plane of the molecule are labeled by $x_i$ and $y_i$, the mass-scaled normal mode coordinates are given by~\cite{schumacher78}
\begin{equation}
         \label{eq:ncoor}
         \begin{array}{l}
         Q_s= \frac{\sqrt{m}}{\sqrt{3}}\left\{-x_1+\left(\frac{1}{2}x_2-\frac{\sqrt{3}}{2}y_2\right)+\left(\frac{1}{2}x_3+\frac{\sqrt{3}}{2}y_3\right)\right\},\\ \\
                  Q_x= \frac{\sqrt{m}}{\sqrt{3}}\left\{-x_1+\left(\frac{1}{2}x_2+\frac{\sqrt{3}}{2}y_2\right)+\left(\frac{1}{2}x_3-\frac{\sqrt{3}}{2}y_3\right)\right\},\\ \\
                           Q_y= \frac{\sqrt{m}}{\sqrt{3}}\left\{y_1+\left(\frac{\sqrt{3}}{2}x_2-\frac{1}{2}y_2\right)+\left(-\frac{\sqrt{3}}{2}x_3-\frac{1}{2}y_3\right)\right\}.
\end{array}
\end{equation}
It is common to introduce polar coordinates $\left(r,\phi\right)$ for the degenerate $\left(Q_x,Q_y\right)$ coordinates, i.e.
\begin{equation}
\label{eq:polar}
\begin{array}{l}
r=\sqrt{Q_x^2+Q_y^2},\\ \\
\tan\phi=\frac{Q_y}{Q_x}.
\end{array}
\end{equation}

\subsection{Jahn-Teller Hamiltonian}
According to the JT effect, predicted already in 1937~\cite{Jahn-Teller-37}, any non-linear molecule with electronic states degenerate by symmetry will distort from the symmetric configuration causing a lifting of the degeneracy. Consequently, due to the conical intersection of the $E$ state at $D_{3h}$ symmetry, the Li$_3$ molecular system possesses a JT effect. JT models are quasidiabatic representations of the system Hamiltonians for the nuclear motion. The simplest example is the linear (only leading order terms are included) $E\otimes\varepsilon$ JT Hamiltonian given by
 \begin{equation}
  \label{eq:JTH}
  \begin{array}{lll}
  H^d & = & T+{\bf U}^d=T+V_{0a}+\frac{1}{2}V_{2a}(Q_x^2+Q_y^2)\\ \\
  & & +V_{1e}\left[\begin{array}{c c} 0 &Q_x-iQ_y\\Q_x+iQ_y& 0 \end{array}\right].
  \end{array}
   \end{equation}  
Here, $V_{0a}$ is the energy of the intersection, $V_{2a}$ is the force constant and $V_{1e}$ is the linear JT coupling parameter.
This model provides a reliable description of the system close to the conical intersection where only the leading terms dominate. To include the nuclear motion in the totally symmetric normal mode coordinate $Q_s$, the JT parameters $V_{2a}$ and $V_{1e}$ are allowed to vary with $Q_s$. Naturally, this will cause a correlation between the nuclear dynamics in the $\varepsilon$ mode and the motion in the totally symmetric mode. As will be shown, we here have parameters of the Hamiltonian that depend smoothly on $Q_s$. 

For obtaining a more reliable description further away from the point of degeneracy,
higher order terms must be included. A general factorized expression of the diabatic potential energy matrix is given by~\cite{viel04} 
\begin{equation}
\label{eq:jthp} 
{\bf U}^d = \sum_{n}\frac{1}{n!}\left[ \begin{array}{c c} V^{(n)}& W^{(n)}-iZ^{(n)} \\ W^{(n)}+iZ^{(n)} & V^{(n)} \end{array}\right]. 
\end{equation}
Here, the matrix elements $V^{(n)}$,$W^{(n)}$, and $Z^{(n)}$ are real functions of the nuclear coordinates $Q_x$ and $Q_y$. The diagonal elements $V^{(n)}$ correspond to the degenerate diabatic potential in the absence of the  JT coupling, while  $W^{(n)}$ and $Z^{(n)}$ are the off-diagonal coupling elements of order $n$. In this work, terms up to third order have been taken into account~\cite{viel04}
  \[\begin{array}{l}
   V^{(0)}= V_{0a},\\
   V^{(1)}= 0,\\
   V^{(2)}= V_{2a}[Q_x^2+Q_y^2],\\
   V^{(3)}= V_{3a}[2Q_x^3-6Q_x Q_y^2],\\
   \\
   W^{(0)}= 0,\\
   W^{(1)}= V_{1e}Q_x,\\
   W^{(2)}= V_{2e}[Q_x^2-Q_y^2],\\
   W^{(3)}= V_{3e}[Q_x^3+Q_x Q_y^2],\\
   \\
   Z^{(0)}= 0,\\
   Z^{(1)}= V_{1e}Q_y,\\
   Z^{(2)}= -2V_{2e}Q_x Q_y,\\
   Z^{(3)}= V_{3e}[Q_x^2Q_y+Q_y^3].\\
	\end{array}
	  \]
The anharmonicity in the $\varepsilon$-mode is contained in the $V_{3a}$ term, while the linear, quadratic, and cubic coupling parameters are represented by $V_{1e}$, $V_{2e}$, and $V_{3e}$ respectively. Again, the parameters of the Hamiltonian are functions of $Q_s$. With the explicit forms of the coupling terms, the elements of the diabatic potential matrix take the form
\begin{equation}\label{eq:ud}
\begin{array}{l}
U^d_{11}=U^d_{22}=V_{0a}+V_{2a}r^2+V_{3a}r^3\cos3\phi\\ \\
U^d_{12}=U^{d*}_{21}=V_{1e} r e^{-i\phi}+V_{2e}r^2e^{2i\phi}+V_{3e}r^3e^{-i\phi}
\end{array}
\end{equation}
when expressed in polar coordinates.

The diagonalization of the diabatic potential energy matrix [eq. (\ref{eq:a2d})] yields the adiabatic potential energy surfaces as eigenvalues.
The corresponding adiabatic potential energy surfaces are given by~\cite{Ernst-10}
\begin{equation}
\label{eq:adpes}
U^a_{1,2}=V_{0a}+V_{2a}r^2+V_{3a}\cos\left(3\phi\right)r^3\pm r\sqrt{V\left(r,\phi\right)},
\end{equation}
where
\begin{equation*}
\begin{array}{lll}
V\left(r,\phi\right) & = & V_{1e}^2+2V_{1e}V_{2e}\cos\left(3\phi\right)r+\left(2V_{1e}V_{3e}+V_{2e}^2\right)r^2\\ \\
& & +2V_{2e}V_{3e}\cos\left(3\phi\right)r^3+V_{3e}^2r^4.
\end{array}
\end{equation*}
The transformation matrix ${\bf S}$ used to transform from the diabatic to adiabatic representations has in general the form
\begin{equation}
{\bf S} = \frac{e^{-is\gamma}}{\sqrt{2}}\left[\begin{array}{c c}
    1 &1\\
    e^{i\gamma}& -e^{i\gamma}
\end{array} 
\right]. 
\label{eq:trsm}
\end{equation}
Here, the constant $s$ is deeply related to the underlying gauge structure of the system. As discussed above, it can be seen as a gauge choice and we will especially pick $s=\frac{1}{2}$ \cite{koppel99,Koppel-89}. With $s=1/2$ we note that the transformation matrix is not singled valued under a rotation $\gamma:\,0\rightarrow2\pi$ which reflects the $\pi$ Berry phase connected with the conical intersection~\cite{baer06}. Including terms up to cubic order in the JT Hamiltonian, the angle $\gamma$ which characterizes the transformation matrix takes the form
\[
\tan{\gamma}=\frac{V_{1e}r\sin\phi-V_{2e}r^2\sin2\phi+V_{3e}r^3\sin\phi}{V_{1e}r\cos\phi+ V_{2e}r^2\cos2\phi+V_{3e}r^3\cos\phi}.
\]

In the diabatization procedure suggested by Thiel and K\"{o}ppel~\cite{koppel99}, the {\it ab initio} adiabatic potential energy surfaces of the two lowest electronic states of Li$_3$ are in the vicinity of the conical intersection fitted to the eigenvalues of the JT Hamiltonian. As described in section~\ref{sec:diab}, the order of the JT Hamiltonian is successively increased until satisfactory convergence of the fitting of the adiabatic potential energy surfaces is obtained. With the fitted parameters, the transformation matrix ${\bf S}$ is then constructed and the computed adiabatic surfaces are transformed to obtain the diabatic potential energy matrix over the whole range of coordinates where the potentials are computed. This is an effective and reliable diabatization procedure that without direct computation of non-adiabatic couplings removes the singular parts of the  couplings~\cite{koppel99}. Far from the conical intersection the adiabatic states are assumed not to interact and the asymptotic form of the transformation matrix is used. As is often the case, also in the Li$_3$ trimer, additional conical intersections appear for large internuclear distances. It should be remembered that the present method fails to give an accurate description at these distances.

\section{Quantum chemistry calculations}\label{sec3}
In this section we present the results from the quantum chemistry computations. All our calculations are performed using the MOLPRO program package~\cite{molpromanual}.

\subsection{Potential energy curves of Li$_2$}
\begin{figure}[htbp]
	\centering
	\includegraphics[width=10cm]{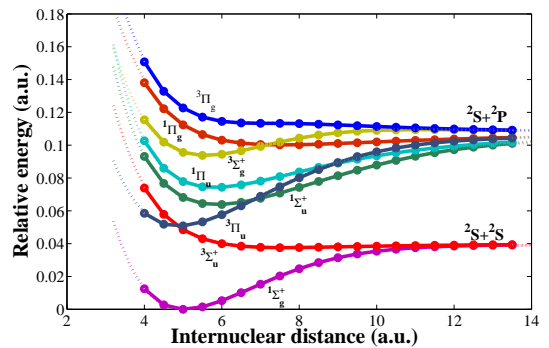}
	\caption {Potential energy curves of Li$_2$ correlated with the Li($^2S$)+Li($^2S$) and Li($^2S$)+Li($^2P$) asymptotic limits. The solid curves with symbols are the potential energy curves computed here, while the dashed curves show the results of Jasik \textit{et al.}~\cite{Jasik-06}.}	
	\label{fig:pecjasic}
\end{figure}

To confirm the accuracy of the quantum chemistry calculations on Li$_3$, we start by performing calculations on the diatomic Li$_2$ system at the same level of theory. We have computed all molecular states of Li$_2$ associated with the two lowest dissociation limits, Li($^2S$)+Li($^2S$) and Li($^2S$)+Li($^2P$). We use the aug-cc-PVTZ basis set following Dunning \textit{et al.}~\cite{Dunning-89}. This is composed of the ($12s,6p,3d,2f$) primitive basis functions contracted into ($5s,4p,3d,2f$). Using SCF (self consistent field) molecular orbitals of the groundstate of Li$_2$, a state-averaged MCSCF (multi-configuration self consistent field) calculation is performed using an active space consisting of all 8 orbitals composed by the $2s$ and $2p$ atomic orbitals. The $1\sigma_g$ and $1\sigma_u$ core orbitals are kept doubly occupied in the MCSCF calculation. 
We then perform a MRCI (multi-reference configuration interaction) calculation using the same active space as for the MCSCF calculation. Furthermore, excitations out of the core orbitals as well as single and double external excitations are included in the CI (configuration interaction) wave function. The potential energy curves calculated for internuclear distances ranging from 4-14 a.u. are displayed in Fig.~\ref{fig:pecjasic}. We compare our potential energy curves with those calculated by Jasik \textit{et al.} using a CI technique~\cite{Jasik-06}, where only the valence electrons are treated explicitly. As can be seen an excellent agreement is obtained. We also achieve very good agreement with the potential energy curves reported by Hotta \textit{et al.} \cite{Hotta-02}.

We have performed MCSCF/MRCI calculations using larger active spaces (up to 18 molecular orbitals have been considered). The shape of the computed potential energy curves are very similar to those presented here. We thus conclude that a MCSCF/MRCI calculation including an active space composed of the molecular orbitals formed by the $2s,2p$ atomic orbitals produce satisfactory potential energy curves for the lower electronic states. With this knowledge, the same active space is also used for our quantum chemistry calculations on Li$_3$.
  
\subsection{Adiabatic potential energy surfaces of Li$_3$}
Starting with a SCF calculation on the ground state of Li$_3$ with the aug-cc-PVTZ basis set~\cite{Dunning-89}, the molecular orbitals are generated and used in the subsequent MCSCF calculation. The active space in the MCSCF calculation is composed of the three valence electrons distributed among all the twelve molecular orbitals composed by the $2s$ and $2p$ atomic orbitals. A state averaged calculation is perform where equal weights are used for the electronic states computed. In the following MRCI calculations, excitations out of the three core orbitals as well as single external excitations are included.

\begin{figure}[htbp]
	\centering
	\includegraphics[width=8cm]{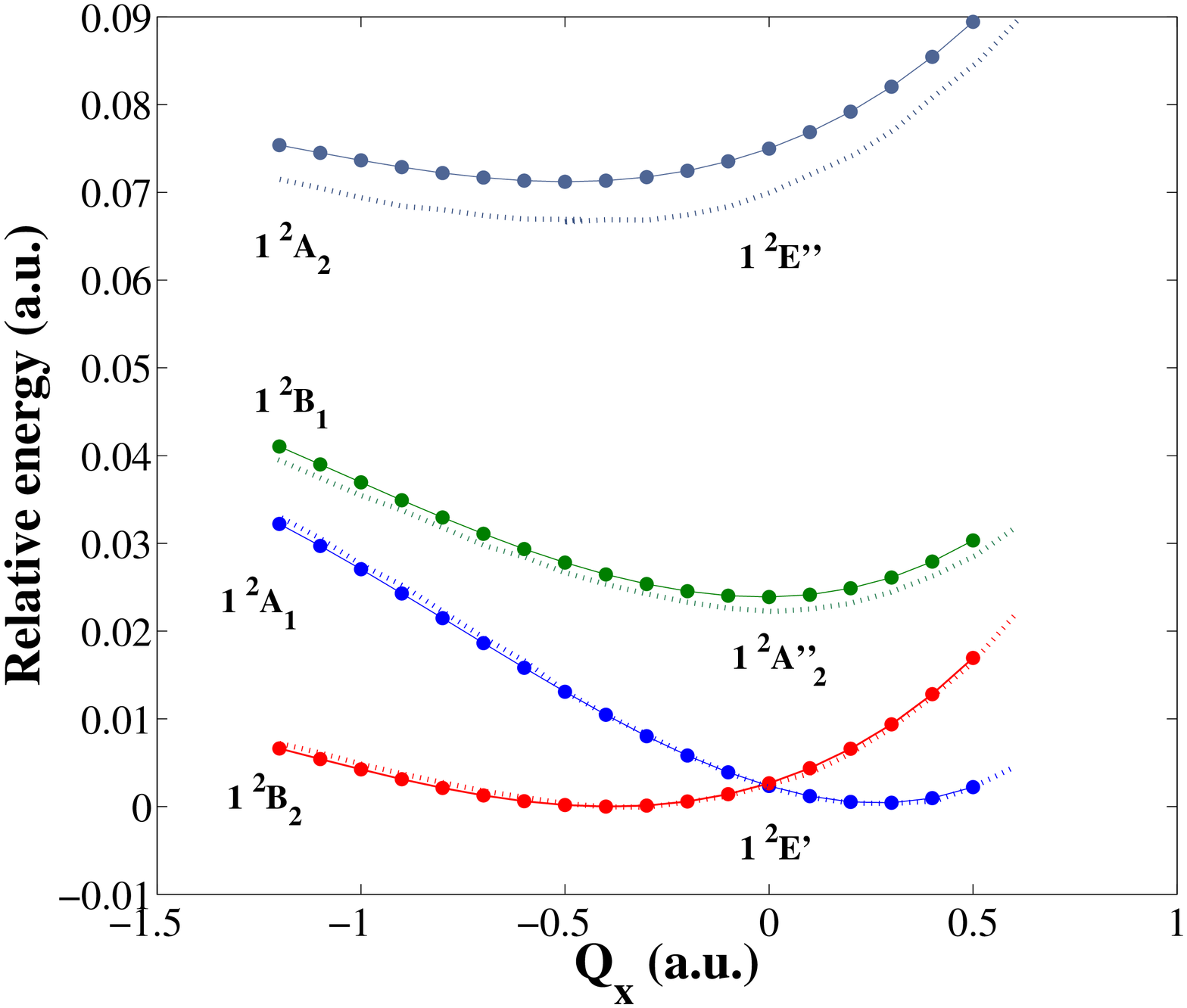}
	\caption{Adiabatic potential energy surfaces of the four lowest states of Li$_3$ for fixed normal mode coordinates $Q_y=0.0$ a.u. and $Q_s= 3.2$ a.u. The dashed curves are the potential energy surfaces computed by Ehara \textit{et al.} \cite{ehara99}.}	
	\label{fig:Li3curves}
\end{figure}

We compute the potential energy surfaces of the lowest four electronic states of Li$_3$. Fixing the normal mode coordinates $Q_y=0.0$ a.u. and $Q_s=3.2$ a.u. (the equilibrium values for the ground state), the potential energies are calculated in $C_{2v}$ symmetry, when $Q_x$ varies from $-1.2$ a.u. to $0.6$ a.u. The computed potential energy surfaces are displayed in Fig.~\ref{fig:Li3curves} with symbols connected by filled lines. 
The {\it ab initio} calculations reveal a global minimum on the $1^2B_2$ state (in $C_{2v}$ symmetry) at the obtuse geometry $Q_s=3.2$ a.u. and $Q_x=-0.4$ a.u.. This corresponds to a bond length of $2.77$ {\AA} and an angle of $73^\circ$. The potential energy surface of the $1^2A_1$ electronic state has a local minima at the acute geometry $Q_s=3.2$ a.u. and $Q_x=0.3$ a.u., corresponding to a bond length of $3.08$ {\AA} and a bond angle of $51^\circ$. The minima of the two lowest electronic states are separated by the conical intersection at the totally symmetric configuration. When the symmetric stretch normal mode coordinate $Q_s$ is varied, a conical intersection seam at $Q_x=Q_y=0$ a.u. is formed.

The dashed lines in Fig.~\ref{fig:Li3curves} show the potential energy surfaces of the four lowest states of Li$_3$ computed by Ehara \textit{et al.} \cite{ehara99} using a MCSCF/MRCI approach with a similar active space in the MCSCF calculation as the one used by us, but with a smaller reference space in the subsequent MRCI calculation. The potential energy surfaces displayed in Fig.~\ref{fig:Li3curves} are shifted in energy to obtain zero at the equilibrium structure. The energy of the equilibrium structure is $E_0=-22.40196023$ H in our calculation, while Ehara and co-workers reported a value of $E_0=-22.357736$ H. We obtain a very good agreement with the potential energy surfaces calculated by Ehara for the lowest two electronic states of Li$_3$. Our excitation energies for the $1^2B_1$ and $1^2A_2$ are larger than the ones obtained by Ehara {\it et al.}~\cite{ehara99}.

\begin{figure}[h]
	\centering
	\includegraphics[width=8cm]{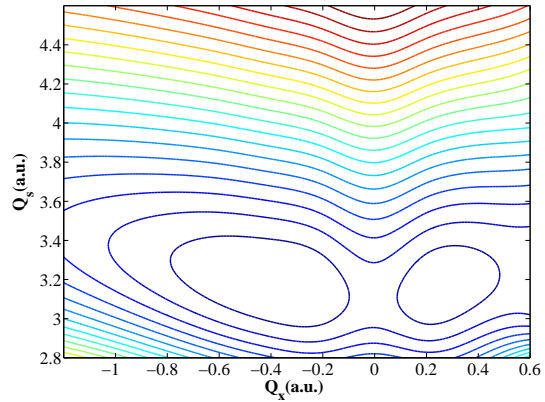}
	\caption {Contour plot of the lowest potential energy surface of Li$_3$ at the fixed asymmetric normal mode coordinates, $Q_y= 0.0$ a.u.}	
	\label{fig:Li3contour}
\end{figure}

A contour plot of the lowest adiabatic potential energy surface is displayed in Fig.~\ref{fig:Li3contour}, where the asymmetric stretching normal mode coordinate $Q_y$ is fixed at zero (keeping $C_{2v}$ symmetry). The symmetric stretching coordinate $Q_s$ is varied between 2.8 and 4.6 a.u. and the bending coordinate $Q_x$ from -1.2 to 0.6 a.u. The discrete \textit{ab initio} results are interpolated with cubic splines to provide a smooth surface.
Two minima can be found on the surface, separated by a sharp cusp at the conical intersection between the lowest two electronic states. 
 
\begin{figure}[htbp]
	\centering
\includegraphics[width=13cm]{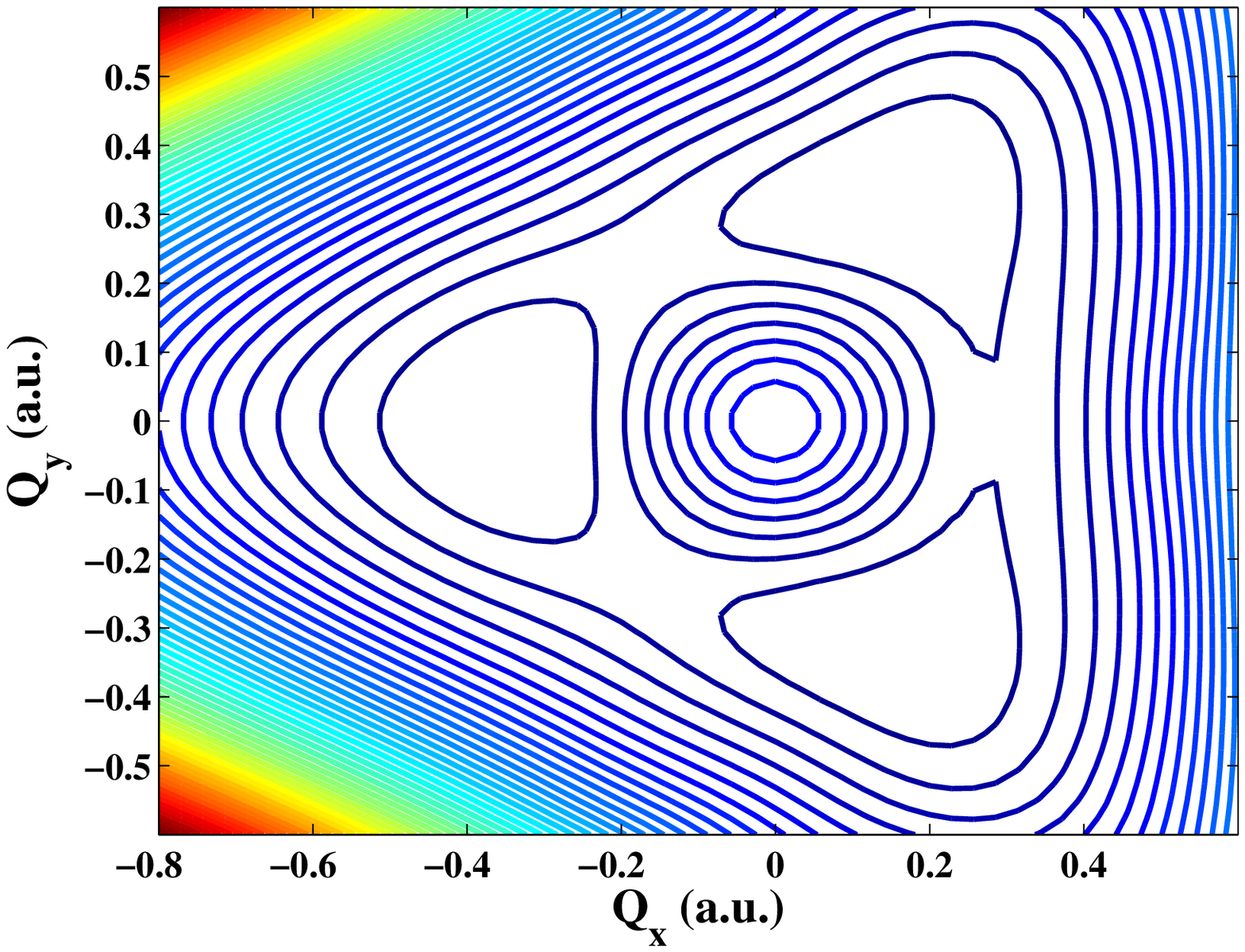}
\caption {Contour plot of the lowest adiabatic potential energy surface of the Li$_3$ at fixed $Q_s = 3.2$ a.u.}
  \label{fig:conplot}	
\end{figure}

When $Q_y$ is nonzero, the $C_{2v}$ symmetry is broken and the system has $C_s$ symmetry. Fig.~\ref{fig:conplot} shows the contour plot of the lowest adiabatic potential energy surface of Li$_3$ when the symmetric normal mode coordinate $Q_s$ is fixed at the equilibrium value 3.2 a.u.. 
The potential energy surface exhibit three minima at the polar coordinates [see eq. (\ref{eq:polar})] $r_m= 0.4$ a.u. and $\phi=\frac{\pi}{3},\pi,\frac{5\pi}{3}$. The minima are separated by barriers toward pseudorotation. The heights of the barriers are $0.000440$ H. The conical intersection located at the totally symmetric configuration has an energy of $0.00235$ H relative to the minima of the potential. The heights of the barrier and the relative energy of conical intersection to the minima of the potential were reported as $0.000337$ H and $0.00231$ H, respectively by Ehara {\it et al.}~\cite{ehara99}. Meyer and co-workers~\cite{Demtroder-2000} found barrier heights of $0.000329$ H and an energy of the conical intersection of $0.00229$ H also by performing {\it ab initio} MRCI calculation.

The lowest two adiabatic potential energy surfaces of Li$_3$ are calculated as functions of the three normal mode coordinates. The symmetric stretch coordinate $Q_s$ is varied between 2.8 a.u. and 4.6 a.u. in steps of 0.1 a.u. The symmetry-lifting coordinates $Q_x$ and $Q_y$ are varied from -10.0 to +10.0 a.u. The quantum chemistry calculations are carried out using polar coordinates. For the radial coordinate $r$, a dense grid with $dr=0.01$ a.u. is used close to the conical intersection when $0.0$ a.u.$\leq r \leq 0.1$ a.u. In $0.1$ a.u.$\leq r \leq 2.0$ a.u. a step-size of $dr=0.2$ a.u. is used while for $2.0$ a.u.$\leq r \leq 10.0$ a.u. $dr=1.0$ a.u. is applied. For the angular coordinate the symmetry of the problem is used and the potential energies are computed between $0 \leq \phi \leq \frac{2\pi}{3}$ with $d\phi=\frac{\pi}{36}$. 

\begin{figure}[htbp]
	\centering
	\includegraphics[width=15cm]{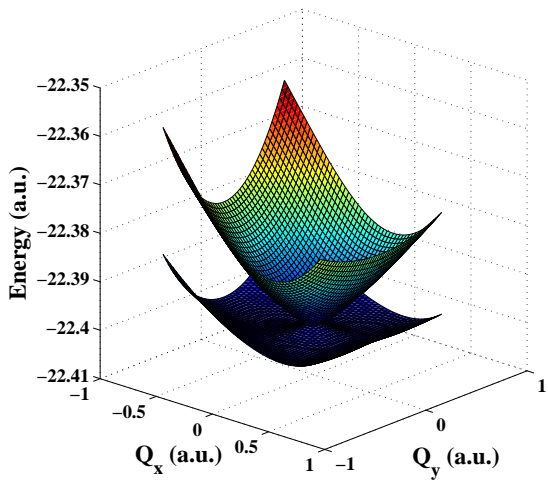}
	\caption {Lowest two adiabatic potential energy surfaces of Li$_3$ at fixed symmetric normal mode coordinate $Q_s = 3.2$ a.u.}	
	\label{fig:pes4}
\end{figure}

Fig.~\ref{fig:pes4} displays the lower two adiabatic potential energy surfaces of Li$_3$ for fixed  $Q_s=3.2$ a.u. as functions of $Q_x$ and $Q_y$. The conical intersection between the two surfaces at $Q_x=Q_y=0$ a.u. can be seen. Further away from the conical intersection (not shown in the figure) three additional conical intersections appear~\cite{baer06} for $r\sim4$ a.u..

\section{Diabatization}
\label{sec:diab}
To perform the diabatization, the lowest two computed adiabatic potential energy surfaces of Li$_3$ are first fitted to the $E\otimes\varepsilon$ JT Hamiltonian close to the conical intersection. The parameters of the Hamiltonian are used to construct the transformation matrix (\ref{eq:trsm}) and the full adiabatic surfaces are then transformed to a diabatic representation.

\subsection{Fitting to the Jahn-Teller Hamiltonian}
The \textit{ab initio} data of the two lowest adiabatic potential energy surfaces are fitted to the eigenvalues of a JT Hamiltonian using
a least-square method. The fitting is performed relative close to the conical intersection where the JT Hamiltonian can be assumed to be valid. More precisely, here we have performed the fitting of the data within $Q_x,Q_y\in\left[-1.0\text{ a.u.},+1.0\text{ a.u.}\right]$. Within this range the outer three conical intersections at $r\sim4$ a.u. should not be of great importance. The fitting is performed for all values of the symmetric normal mode coordinate $Q_s$, and the JT parameters are thus given as functions of $Q_s$.

We have used different models of the JT Hamiltonian including an increasingly number of parameters. The simplest ``linear" JT Hamiltonian (here referred to as JT$_1$) describes the diabatic potential energy surfaces as harmonic oscillators and the coupling between the two states is given as a linear function of the symmetry-breaking coordinates $Q_x$ and $Q_y$. This is the Hamiltonian given by eq. (\ref{eq:JTH}). The corresponding lowest adiabatic potential energy surface has a ``Mexican hat" shape with a cylindrical symmetry, i.e. no dependence of the polar angle $\phi$. 

When the quadratic term of the coupling is included (JT$_2$), the adiabatic surfaces become warped and now three global minima of the lowest surface develop.
Adding more terms to the JT model such as the anharmonicity of the diabatic potential energy surfaces and cubic coupling terms, results in additional terms that will break the cylindrical symmetry. The models where these terms added are labeled JT$_3$-JT$_6$. According to eq. (\ref{eq:adpes}), the potential energy surface of the lowest adiabatic state will obtain either minima or maxima (barriers) at $\phi=\frac{\pi}{3},\pi,\frac{5\pi}{3}$ depending on the relative magnitudes of the anharmonicity and the coupling terms.  

The optimized parameters of the different JT Hamiltonians JT$_1$ - JT$_6$ are given in Table.~\ref{tab:fitres}.
\begin{table}
	\centering
		\begin{tabular}{c c c c  c c c}
		\hline \hline
		JT-model    & $V_{0a}$     &$V_{2a}$ & $V_{3a}$&$V_{1e}$  & $V_{2e}$& $V_{3e}$\\ 
		             & [$a.u.$]&[$a.u.$]&[$a.u.$]&[$a.u.$]&[$a.u.$]&[$a.u.$]\\ \hline 
		$JT_1$   &-22.3994415 &0.022& -    &0.0140&-&-\\
		$JT_2$   &-22.3994415 &0.022& -    &0.0140&0.0012&-\\
		$JT_3$   &-22.3994415 &0.022& -    &0.0140&0.0036&-0.0016\\
		$JT_4$  &-22.3994415 &0.022&0.0098&0.0140&-&-\\
		$JT_5$  &-22.3994415 &0.022&0.0095&0.0140&0.001&-\\
		$JT_6$  &-22.3994415 &0.0229&0.0098&0.0142&-0.0013&-0.0014\\ \hline \hline    
	
	\end{tabular}	\caption{Parameters obtained from fitting the lowest two adiabatic potential energy surfaces to the different JT models at $Q_s=3.2$ a.u..}
\label{tab:fitres}
\end{table}
It can be noted that adding higher order terms to the Hamiltonian, does not significantly change the magnitudes of the terms of lower order.
 
It is common to use scaled parameters~\cite{scaledJT} of the JT Hamiltonian. The obtained scaled parameters obtained by fitting of the surfaces at $Q_s=3.2$ a.u. are compared with published values in Table~\ref{tab:jtpar}. 
\begin{table}
\centering
\begin{tabular}{c c c c c c }
		\hline \hline
  system & $k$         &  $g$         &$\omega_e$  &  Ref \\ 
         &             &              & [$cm^{-1}$]&    \\ \hline 
  Li$_3$ &2.007        &0.227        &245.06      &    present work     \\
         &1.96$\pm$0.33&0.22$\pm$0.07 &278$\pm$61  &     \cite{Bersuker-06, Peyerimhoff-84}\\
         &2.25$\pm$0.24&0.14$\pm$0.06 &250$\pm$41  &     \cite{Bersuker-06, martin83}\\ \hline \hline
  \end{tabular} 
  \caption{Scaled parameters of the JT Hamiltonian at $Q_s=3.2$ a.u. are compared with other theoretical predictions.} 
 \label{tab:jtpar}
\end{table}

\begin{figure}[htbp]
	\centering
	\includegraphics[width=9cm]{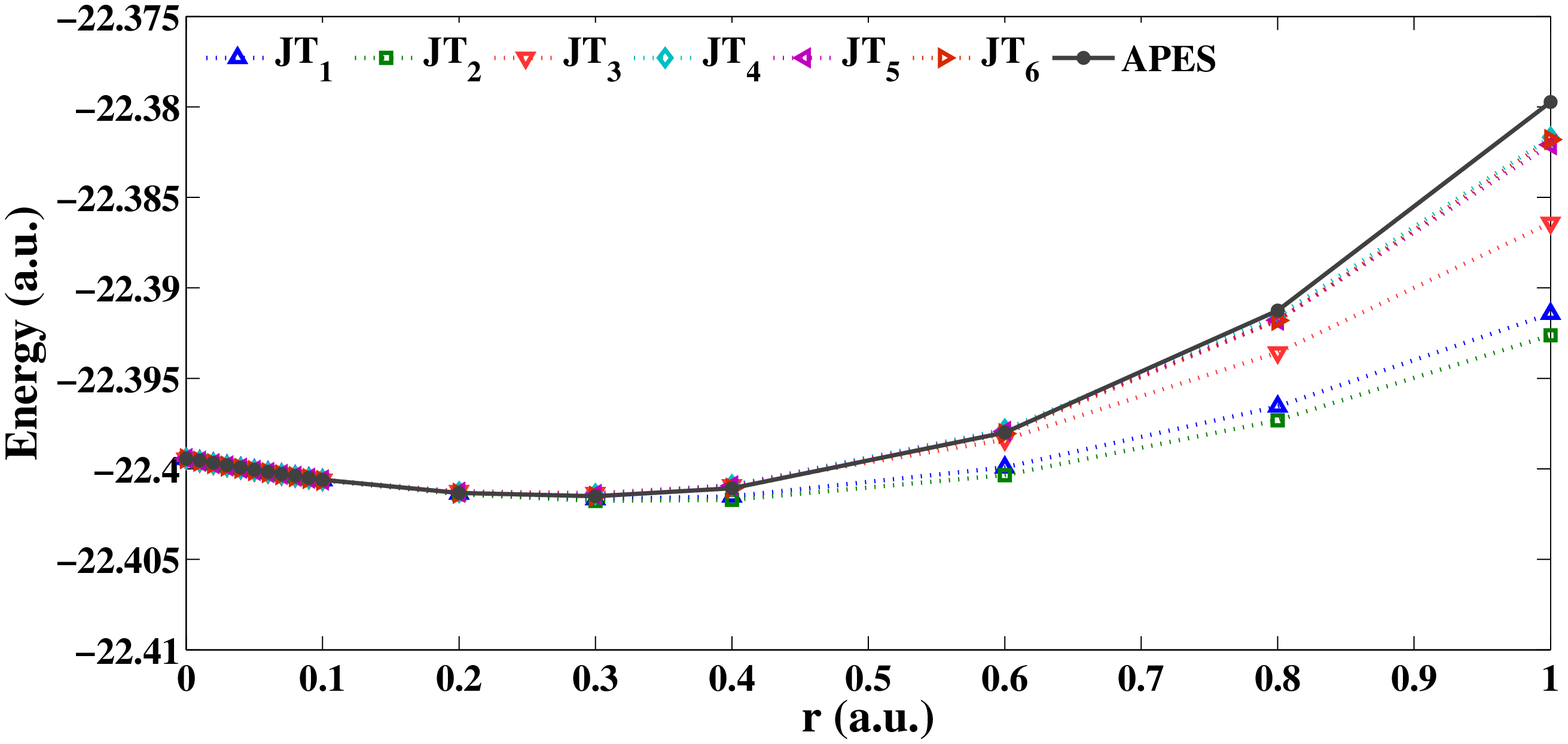}
	\caption {The lower adiabatic potential energy surface along the $r$  polar coordinate for fixed $\phi= 0$ and $Q_s=3.2$ a.u.. } 	
	\label{fig:fittr-03}
\end{figure}

\begin{figure}[htbp]
	\centering
	\includegraphics[width=9cm]{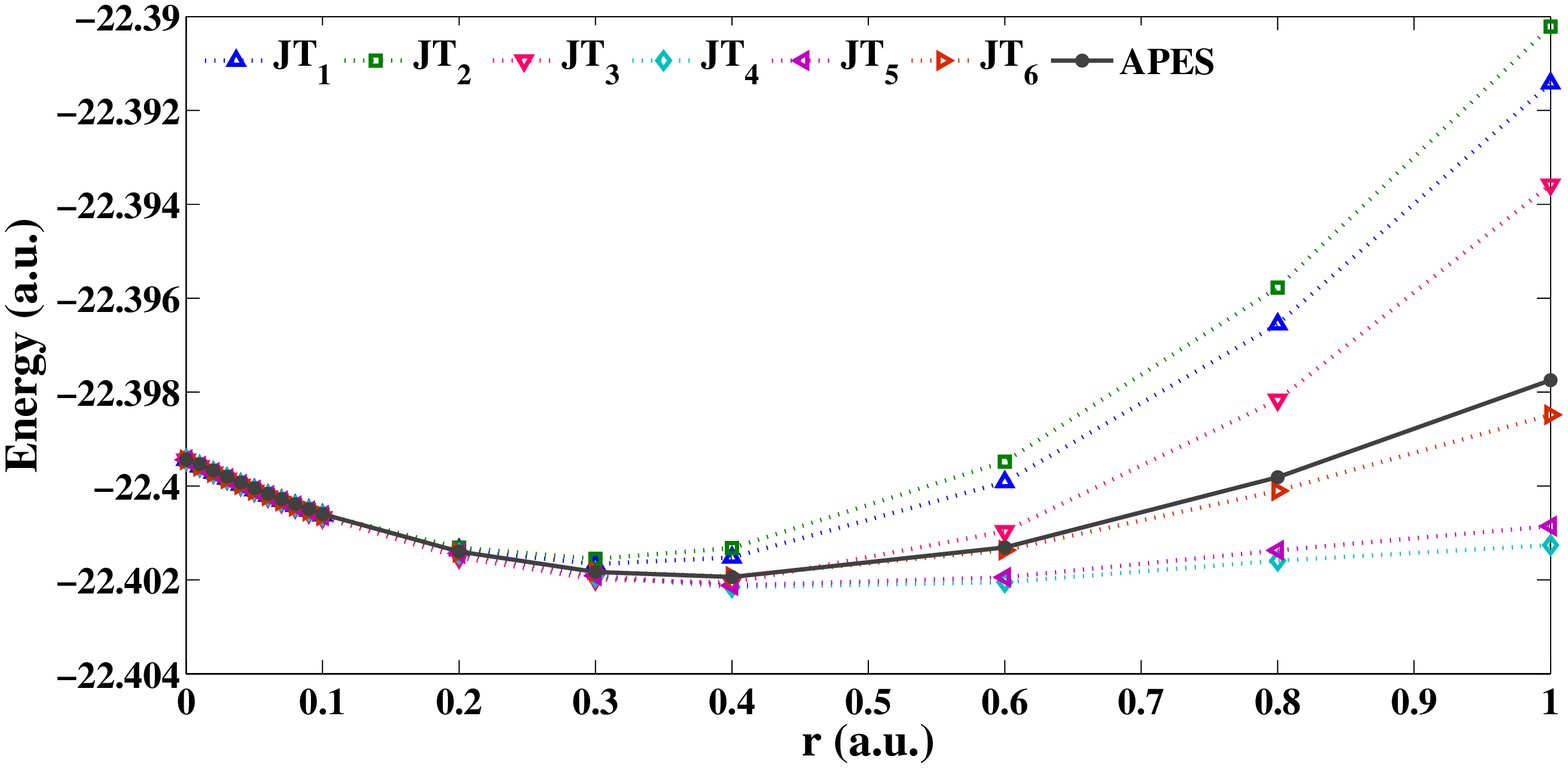}
	\caption { Same as Fig.~\ref{fig:fittr-03} but for fixed $\phi=\frac{\pi}{3}$.  }	
	\label{fig:fittr-603}
\end{figure}

In Fig.~\ref{fig:fittr-03} and Fig.~\ref{fig:fittr-603}, the fits of the lowest adiabatic potential energy surface at $\phi=0$ and $\phi=\frac{\pi}{3}$ are displayed with open symbols connected by dotted lines. The filled symbols are the results from the {\it ab initio} calculations. As expected, further away from the conical intersection the agreement is improved when higher order terms are included in the JT expansion.

\begin{figure}[htbp]
	\centering
	\includegraphics[width=9cm]{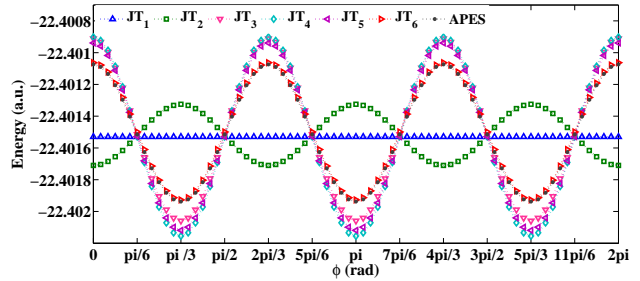}
	\caption {The lower adiabatic potential energy surfaces along the $\phi$ polar coordinate for fixed $r=r_m$, where $r_m$ is the distance from the conical intersection to the minimum of the potential energy surface. }	
	\label{fig:fittphi-rm3}
\end{figure}

In Fig.~\ref{fig:fittphi-rm3}, the results from the fits are compared with the ground potential energy surface as a function of $\phi$ when the polar coordinate is $r_m=0.4$ a.u. (i.e. the distance to the minima of the PES). 
As can be seen, the linear JT$_1$ model provides an adiabatic surface with no dependence on $\phi$. When the quadratic coupling term is included (JT$_2$), the adiabatic surface is warped
and three minima arise separated by barriers. The minima, however, are not formed at $\phi=\frac{\pi}{3},\pi,\frac{5\pi}{3}$ where the {\it ab initio} calculations predict them to be. Rather they are shifted by $\frac{\pi}{3}$.
When the anharmonicity term, or higher order coupling terms are added to the Hamiltonian, the minima are obtained at the right polar angles. 
Here it is the anharmonicity term $V_{3a}$ causing the warping of the potential energy  surface. The fits of the potential energy surfaces in the JT$_6$ model including up to cubic terms both on the diagonal and off-diagonal elements, produce reliable fits for $r\leq1.0$ a.u. For larger polar distances, the JT Hamiltonian (\ref{eq:JTH}) is no longer valid.

The least-square fits presented above give an estimate how well the JT Hamiltonians reproduce the {\it ab initio} calculated potential energy surfaces. It is clear that the lower order JT Hamiltonian model like $JT_2$ gives a poor representation of the adiabatic potential surfaces. In contrast, the JT Hamiltonian model $JT_6$ is in remarkable agreement with the computed data. 

\subsection{$Q_s$-dependence of the Jahn-Teller parameters}
\begin{figure}[htbp]
	\centering
	\includegraphics[width=11cm]{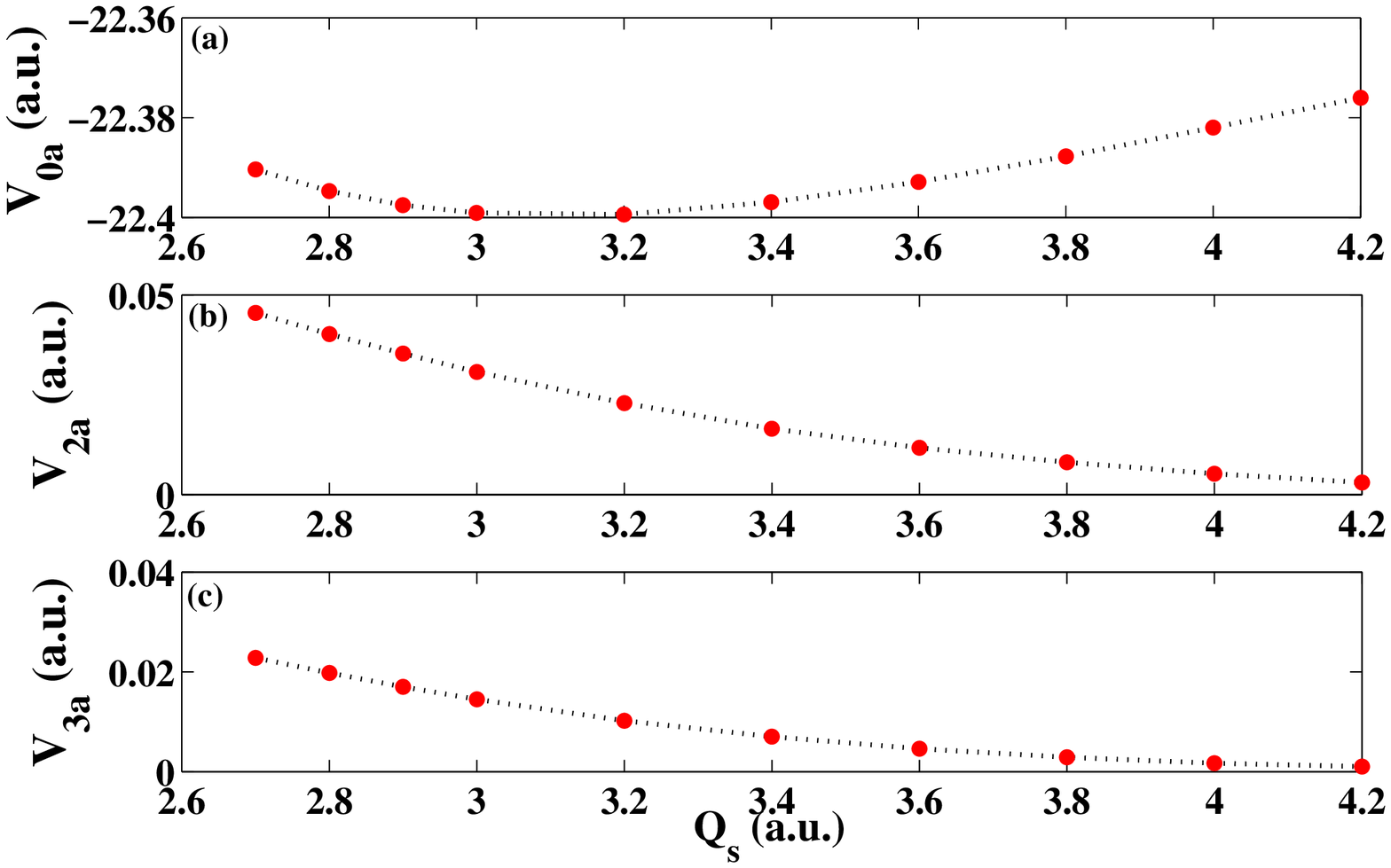}
	\caption {Extracted diagonal JT parameters as functions of the totally symmetric stretching normal mode coordinate $Q_s$. The three plots give respectively, (a) $V_{0a}$, (b) $V_{2a}$, and (c) $V_{3a}$. }	
	\label{fig:jtpar1}
\end{figure}

\begin{figure}[htbp]
	\centering
	\includegraphics[width=13cm]{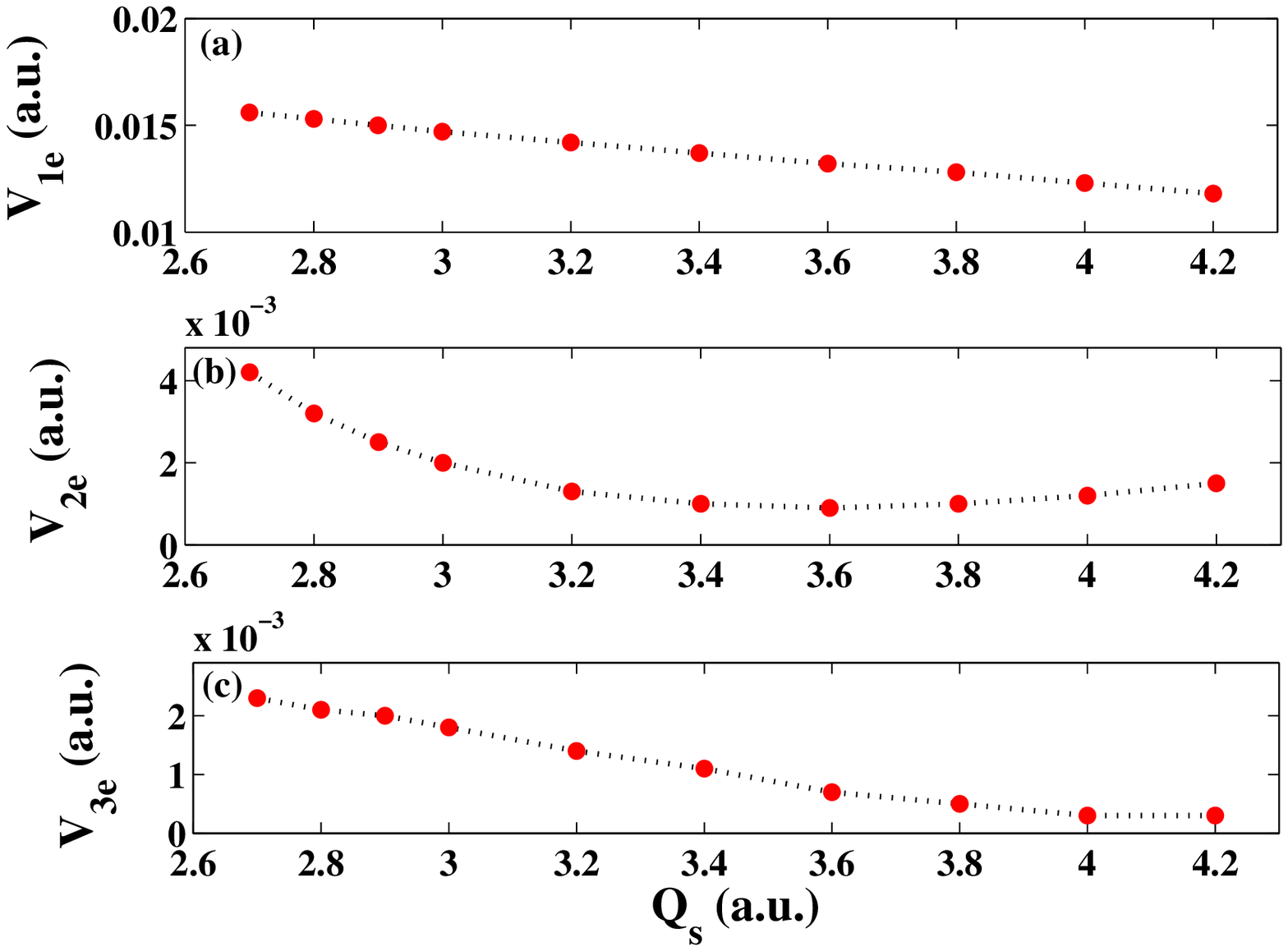}
	\caption {Same as Fig.~\ref{fig:jtpar1} but for the off-diagonal JT parameters. The three plots give, (a) $V_{1e}$, (b) $V_{2e}$, and (c) $V_{3e}$. }	
	\label{fig:jtpar2}
\end{figure}

We extract the JT parameters for each value of the symmetric stretch normal mode coordinate $Q_s$ and investigate the $Q_s$-dependence of these parameters. Figure~\ref{fig:jtpar1} shows the optimized energy of the conical intersection $V_{0a}$, harmonic force constant $V_{2a}$, as well as the cubic anharmonicity $V_{3a}$, as functions of the $Q_s$. Equivalently, in Fig.~\ref{fig:jtpar2}, the linear, $V_{1e}$, quadratic, $V_{2e}$, and cubic, $V_{3e}$, coupling parameters are displayed. For all six parameters the $Q_s$-dependence is smooth, which justifies the picture of the system as a $Q_s$ ``parametrized'' JT model. The energy of the conical intersection, $V_{0a}$, reveals a harmonic dependence of the $Q_s$ coordinate reflecting the symmetric normal mode vibration. We obtain a vibrational frequency of $\omega_a=368.2$ cm$^{-1}$, which should be compared with previous theoretical predictions of $349$ cm$^{-1}$~\cite{schumacher78} and $327$ cm$^{-1}$~\cite{martin83}. 	


			
\subsection{Diabatic potential energy matrix}
\begin{figure}[htbp]
	\centering
  \subfloat
{
\includegraphics[width=0.9 \textwidth]{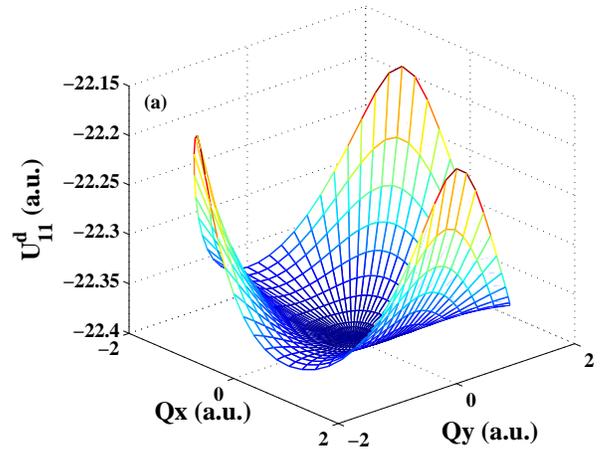}
\label{fig:U_11}
}\\ 
\subfloat
{\includegraphics[width=0.9 \textwidth]{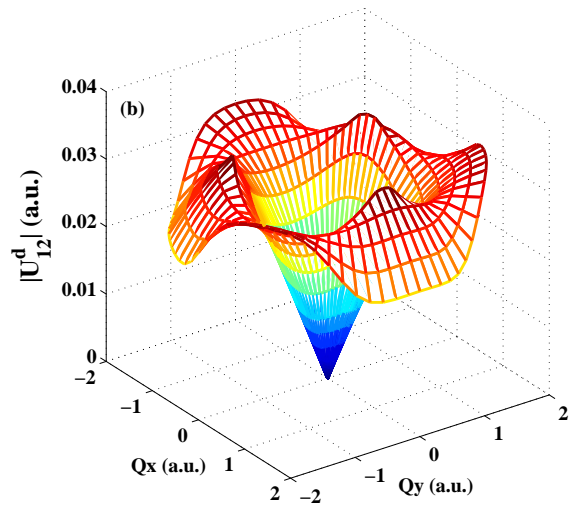}
\label{fig:absU_12}
} \\	
\caption {The elements of the diabatic potential energy matrix, (a) shows the diagonal element $U_{11}^d$ and $U_{22}^d$, while the absolute value of the off-diagonal element $|U_{12}^d|$ is displayed in (b). In both plots $Q_s=3.2$ a.u.. }	
	\label{fig:djt12real}
\end{figure}

By transforming the {\it ab initio} adiabatic potential energy surfaces using the transformation matrix, the diabatic potential energy matrix is constructed. In the region where the JT Hamiltonian is valid, i.e. $r\leq 1.0$ a.u., the two diagonal elements of the diabatic potential matrix are identical [see eq. (\ref{eq:ud})]. The $U_{11}^d$ potential energy surface is displayed in 
Fig.~\ref{fig:U_11} at $Q_s=3.2$ a.u.. The diagonal element is dominated by the harmonic term, however the anharmonicy through the term containing $V_{3a}$ can be seen by the breakdown of the cylindrical symmetry.

Contrary to the diagonal elements, the off-diagonal elements of the diabatic potential energy matrix are complex [see eq. (\ref{eq:ud})]. Fig. ~\ref{fig:absU_12} shows the absolute value of the off-diagonal elements of the diabatic potential energy matrix. The non-zero imaginary part is of great importance as it is related to having a non-trivial Berry curvature that can give rise to anomalous molecular dynamics~\cite{jonas13}. In this plot both the anharmonicity and remnants of the underlying {\it ab initio} adiabatic surfaces are seen in the `melting' structure of the cone for larger radial distances.

\begin{figure}[htbp]
	\centering
	\includegraphics[width=9cm]{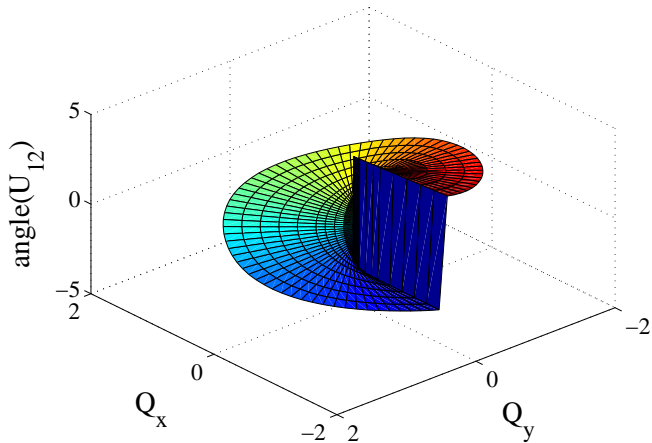}
	\caption {The phase of the electronic coupling term $U^{d}_{12}$ for $Q_s=3.2$ a.u.. }	
	\label{fig:djt12angle}
\end{figure}

In Fig.~\ref{fig:djt12angle}, the phase of the $U_{12}^d$ matrix element is presented. The $2\pi$ phase winding characteristic for the JT system is clearly visible, which also confirms that the JT coupling terms are complex. Furthermore, as discussed in the introduction, it is seen that in the diabatic representation the coupling terms are analytic and free from singularities. Rather the electronic coupling is dominated by the linear coupling term and the effect from the higher order couplings is evident in the fact that the norm is not polar symmetric. 
\begin{figure}[htbp]
	\centering
  \subfloat
{
\includegraphics[width=8cm]{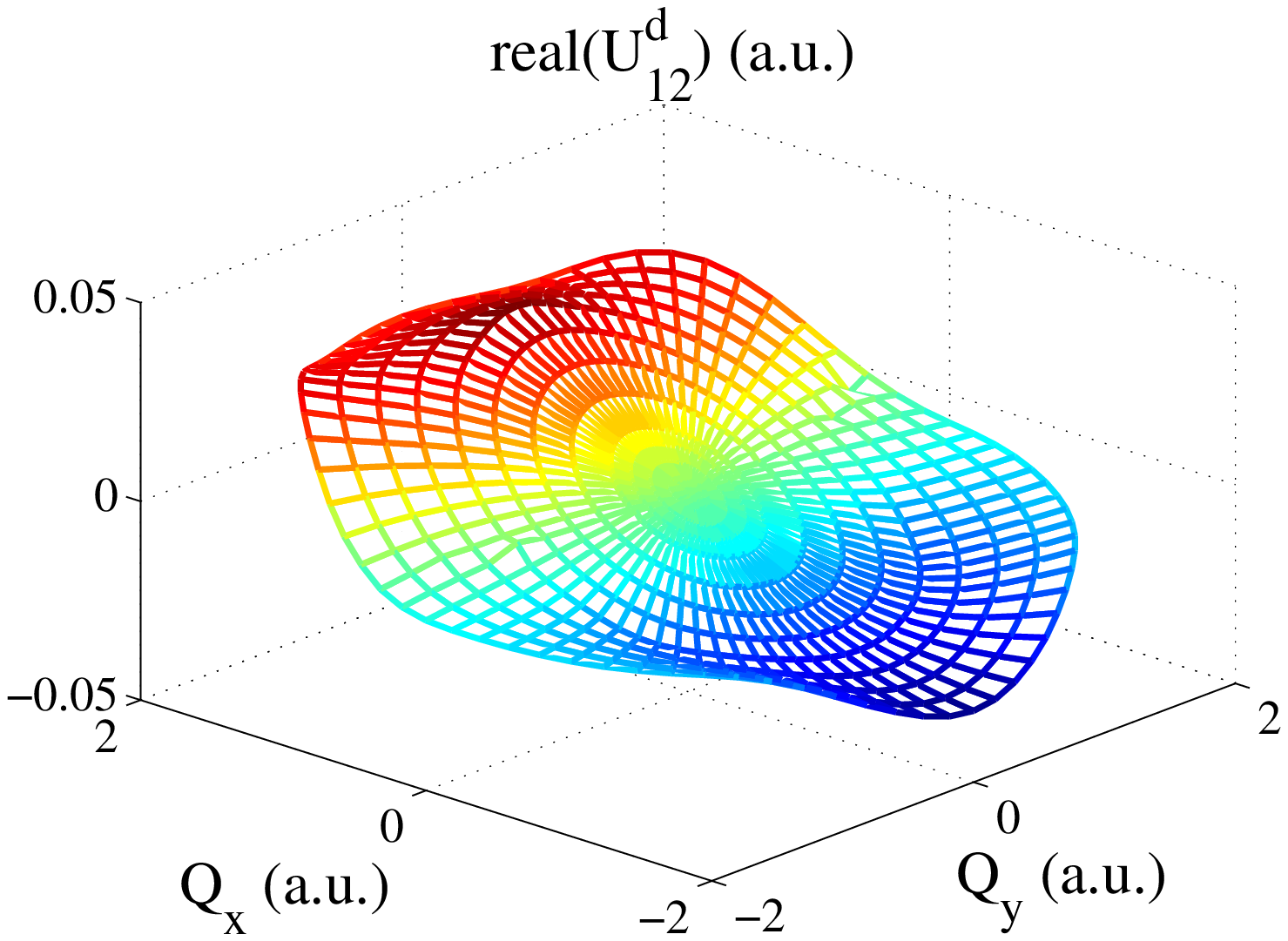}
\label{fig:realU_12}
}\\ 
\subfloat
{\includegraphics[width=8cm]{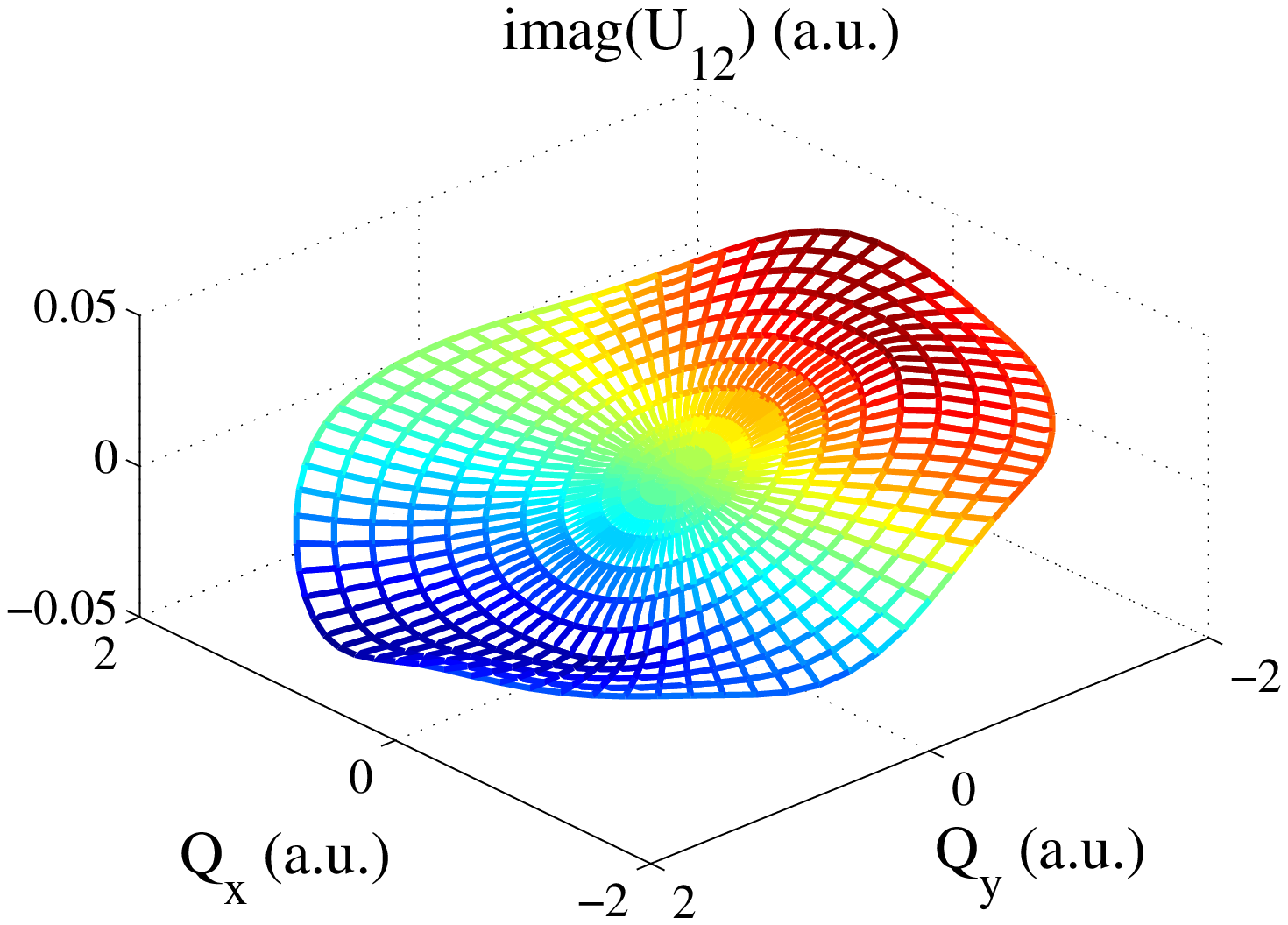}
\label{fig:imagU_12}
} \\	
\caption {The real part (a) and the imaginary part (b) of the electronic coupling term $U^{d}_{12}$ for $Q_s=3.2$ a.u.. }	
	\label{fig:djt12real}
\end{figure}

For completeness, in Fig.~\ref{fig:djt12real} the corresponding real and imaginary parts of the electronic couplings are displayed. As seen in Fig.~\ref{fig:djt12real}, for $r\leq1$ a.u. the linear part is dominating the JT coupling. For larger radial distances the higher order terms become important. Remember that the diabatization method applied here gives a reliable diabatic potential energy matrix even for large inter-nuclear distances as long as the two adiabatic surfaces are well separated. Thus, for $r>3$ when the adiabatic surfaces begin to show signatures of the outer three conical intersections the method is no longer valid. In this respect, if, for example, molecular dynamics is studied with the calculated JT Hamiltonian it must be kept in mind that the radial coordinate $r$ cannot become too large, i.e. molecular vibrations should be kept moderate.

\section{Discussion and conclusion}\label{sec5}
The ground and first excited adiabatic potential energy surfaces of Li$_3$ were computed using the MRCI method. The two adiabatic states exhibit a JT conical intersection at totally symmetric nuclear configurations. Close to the intersection, the surfaces were fitted to the eigenvalues of the potential part of a JT Hamiltonian including terms up to cubic order. We noted that especially the anharmoniciy in the diagonal terms ($V_{3a}$) had to be included in order to correctly describe the polar angular dependence of the adiabatic surfaces. The JT parameters were used to set up a transformation matrix that transforms the {\it ab intio} adiabatic surfaces to a diabatic potential energy matrix. The diabatic representation is reliable not only close to the conical intersection where the JT Hamiltonian is valid but also further out up till the vicinity of the outer three conical intersections that appear in Li$_3$. The computed diabatic potential matrix can  be used to study molecular dynamics in the vicinity of the symmetric conical intersection. Indeed, it is well known that the presence of conical intersections may greatly affect molecular dynamics~\cite{baer06,yarkony} and it has been found that the intersections are crucial for fast radiationless decay of photonexcited organic or biological systems. They have also been found to play in important role in electron recombination processes such as dissociative recombination of polyatomic molecules~\cite{kokoouline01}. As another example, in our previous work~\cite{jonas13} we used the computed diabatic potential energy matrix to illustrate anomalous wave packet dynamics in Li$_3$. Even when the wave packet propagated far from the conical intersection (where the Born-Oppenheimer approximation is supposedly valid) the full coupled system had to be considered in order to correctly explain such anomalous evolution.   

\begin{acknowledgments}
VR (Vetenskapsr\aa det) is acknowledged for financial support.
\end{acknowledgments}
\bibliography{apssamp}

\end{document}